\documentclass[twocolumn,showpacs,preprintnumbers,amsmath,amssymb]{revtex4}

\usepackage{graphicx}
\usepackage{dcolumn}
\usepackage{bm}

\begin{document}

\hyphenation{Es-ta-du-al}

\preprint{IAG-02-2004}

\title{Longitudinal Development of Extensive Air Showers: \\Hybrid Code SENECA and Full Monte Carlo}

\author{Jeferson A. Ortiz}
\email{jortiz@astro.iag.usp.br}
\author{Gustavo Medina-Tanco}
\author{V. de Souza}
\affiliation{Instituto de Astronomia, Geof\'{\i}sica e Ci\^encias Atmosf\'ericas\\ Universidade de S\~ao Paulo,\\ Caixa Postal 9638, S\~ao Paulo, SP 01065-970, Brasil\\}

\begin{abstract}
New experiments, exploring the ultra-high energy tail of the cosmic ray spectrum with unprecedented
detail, are exerting a severe pressure on extensive air shower modeling. Detailed fast codes are in
need in order to extract and understand the richness of information now available. Some hybrid simulation
codes have been proposed recently to this effect (e.g., the combination of the traditional Monte Carlo
scheme and system of cascade equations or pre-simulated air showers). In this context, we explore the potential of
SENECA, an efficient  hybrid tridimensional simulation code, as a valid practical alternative to full Monte
Carlo simulations of extensive air showers generated by ultra-high energy cosmic rays. We extensively
compare hybrid method with the traditional, but time consuming, full Monte Carlo code CORSIKA which is the
de facto standard in the field. The hybrid scheme of the SENECA code is based on the simulation
of each particle with the traditional Monte Carlo method at two steps of the shower development:
the first step predicts the large fluctuations in the very first particle interactions at high
energies while the second step provides a well detailed lateral distribution simulation of the final
stages of the air shower. Both Monte Carlo simulation steps are connected by a cascade equation system
which reproduces correctly the hadronic and electromagnetic longitudinal profile. We study the influence
of this approach on the main longitudinal characteristics of proton-induced air showers and compare the
predictions of the well known CORSIKA code using the QGSJET hadronic interaction model.
\end{abstract}



\maketitle

\section{\label{introduction}Introduction}

\par Since the very first observations, ultra-high energy cosmic rays (UHECR)
have been an open question and a priority in astroparticle physics.
Their origin, nature and possible acceleration mechanisms are still a
mystery. In the last decades many experiments such as Volcano
Ranch~\cite{Linsley61,Linsley63}, Haverah Park~\cite{Lawrence01},
Yakutsk~\cite{Afanasiev01}, Fly's Eye~\cite{Baltrusaitis88,Bird95}, HiRes~\cite{HiRes01a} and
AGASA~\cite{Chiba92,Hayashida94} have contributed for the study of UHECR's,
setting up the existence of such high energy particles. Shorty, the Pierre Auger 
Observatory will begin to explore them in unprecedented detail~\cite{Blumer03,Auger04}.

Due to the very low flux of high energy cosmic rays, measuring extensive air
showers (EAS) is the only possible technique to learn about the shape of the UHECR
spectrum and their chemical composition. Two different ways have been historically
applied to observe and analyze EAS's: ground array of detectors and optical detectors.
Surface detectors measure a lateral density sample and trigger in coincidence
when charged particles pass through them. Optical detectors (i.e., fluorescence
detectors) observe the longitudinal profile evolution by measuring the fluorescence
light from atmospheric nitrogen excitation produced by the ionization of the secondary
charged particles (essentially electrons and positrons). The combination of shower
observables (such as lateral density, the depth of maximum shower development
($X_{\rm max}$) and number of muons ($N_\mu$) at detector observation level) and
simulation techniques is the current way to obtain information about the primary energy,
composition and arrival direction. For this purpose the shower simulation should provide
all possible, and ideally the necessary, information to interpret measurements of shower parameters.
We suggest the reference \cite{Engel04b} which is an interesting summary of experimental 
results from highest energy cosmic ray measurements, focused on data and analyzes that
became available after 1999.

Many modern shower simulation packages have been developed over the years.
Most of them are based on the Monte Carlo method and simulate complete
high energy showers with well described fluctuations in the first particle
interactions and realistic distributions of energy of shower particles.
Unfortunately, the calculation of the gigantic showers induced by cosmic
rays with energies above 10$^{18}$~eV becomes a very difficult technical
problem. This is due to the huge number of created secondary particles that have
to be followed in the Monte Carlo method. As a consequence the direct simulation
of the shower following each individual particle becomes practically impossible
and the computation time takes too long.

Recently, different ways of calculating the air shower development have been
proposed~\cite{Bossard01,Engel04c}. Most of them combine the traditional Monte Carlo
scheme with a system of electromagnetic and hadronic cascade equations. In a
new one dimensional approach~\cite{Alvarez02} pre-simulated pion-induced 
showers are described with parametrizations and then are superimposed 
to pion and kaon particles after their first interaction points are 
simulated by the Monte Carlo method.

In the present work we analyze extensively the results obtained by the
SENECA~\cite{Drescher03} code. The SENECA simulation approach
is based on the Monte Carlo calculation of the first and final stages of the air shower development,
and on a cascade equation system that connects both stages reproducing
the longitudinal shower development. We explore mainly the fast air shower
generation for different primary energies. As an application of this approach
we investigate the main longitudinal shower characteristics of proton,
iron nucleus and gamma initiated air showers up to ultra-high energy, as 
predicted by the QGSJET~\cite{qgsjetb,qgsjet} hadronic interaction model. 
We compare SENECA results  with the well tested CORSIKA 
(COsmic Ray SImulations for KAscade) simulation code~\cite{Heck98a}.

This paper is structured as follows. 
In Sec.~\ref{secII} we briefly describe the hybrid approach and 
the air shower modelling. 
In Sec.~\ref{secIII} the method is applied to study the longitudinal 
development and the shower observables as well. Our main intention is to compare 
our predictions to calculations performed with the CORSIKA shower generator, 
in order to verify the stability of the hybrid code and the reliability of the 
its results, in the sense of being useful to several experimental applications. 
We have made several calculations for gamma, iron nucleus and proton initiated showers with 
different energy thresholds, using the thinning procedure for the Monte Carlo simulation scheme.  
Average and fluctuation values of $X_{\mathrm {max}}$ and $S_{\mathrm {max}}$, 
their correlations and distributions are presented. 
The number of electrons ($N_{e}$) and muons ($N_{\mu}$) have been analyzed at different observation levels. 
Also, we show a comparison between CORSIKA and SENECA 
CPU time requirements for proton showers at different primary energies. 
Section~\ref{secIV} summarizes our results.

\section{\label{secII} SENECA Approach and the air shower modelling}

The main goal of this approach is the generation of EAS's in a
fast manner, obtaining the correct description of the fluctuations
in showers and giving the average values for the shower characteristics.
The detailed technical information about the SENECA package
can be obtained in~\cite{Drescher03,Drescher05}. Even though the SENECA
code describes both longitudinal and lateral air shower developments, the
simulation scheme is used here to generate large statistics of
longitudinal shower profiles applicable mainly to the present fluorescence
detectors, such as Pierre Auger Observatory~\cite{Cronin95} and HiRes~\cite{HiRes01a,HiRes01b}, as well to the future telescope EUSO~\cite{Catalano01}.

For the present work we track explicitly every particle with energy
above the fraction $f$=$E_{0}/1.000$, where $E_{0}$ is the primary shower
energy, studying in detail the initial part of the shower. 
All secondary particles with energy below the mentioned fraction are
taken as initial conditions to initialize a system of hadronic and
electromagnetic cascade equations (see the suggested references). We
use the cascade equations up to several minimum electromagnetic energy thresholds of
1, 3.16, 10, 31.6 and 100~GeV for the electromagnetic component 
($E_{\mathrm {min}}^{\mathrm{em}}$) and $E_{\mathrm {min}}^{\mathrm{had}}$=$10^{4}$~GeV 
for the hadronic component. Hadronic and electromagnetic particles with
energies below the $E_{\mathrm {min}}^{\mathrm{had}}$ and
$E_{\mathrm {min}}^{\mathrm{em}}$ are no longer treated by the
cascade equations but traced again in the Monte Carlo scheme.
The hadronic interactions at high energies are calculated
with the QGSJET01~\cite{qgsjet} model while the interactions at low energies with
GHEISHA~\cite{gheisha}. The electromagnetic interactions are treated by the
EGS4~\cite{egs4} code. The adopted kinetic energy cutoffs for all simulations
were 50~MeV (0.3~MeV) for hadrons and muons (electrons and positrons). 
The forementioned energy thresholds were chosen in order to verify qualitatively
the dependence and performance of the cascade results, trying to minimize 
the CPU time without the introduction of significant errors. 
All SENECA simulations were performed with 1.2.2. version.

\section{\label{secIII}Results and Comparisons}

In this session we apply the SENECA simulation approach to generate gamma,
iron nucleus and proton induced air showers at fixed primary energies and 
explore the longitudinal development. Although the simulation of showers 
at fixed energies is not a very realistic application we intend in the present 
work to compare quantitatively SENECA and CORSIKA results. One important
reason for this comparison is to optimize the compromise between simulation
time usage and accuracy in the description of fluctuations.
In this particular case, such detailed study can be useful to many experiments
which use the fluorescence technique.

\begin{figure}[t]
\centerline{
\includegraphics[width=8.5cm]{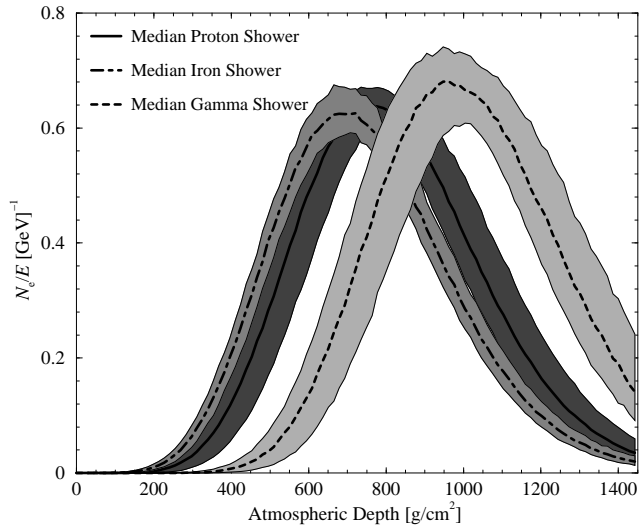}
}
\caption{Median longitudinal profile of electrons-positrons for 10 random 
gamma (dashed), iron nucleus (dot dashed) and proton (solid) induced showers 
at primary energy of 10$^{19}$~eV, with zenith angle $\theta$=$45^{\circ}$, 
calculated by the SENECA scheme, by using the QGSJET01 hadronic interaction 
model. The shaded bands which follow each median line represent the upper limit of 
68\% of confidence level, for 1,000 simulated events.}
\label{long_sen}
\end{figure}

In Fig.~\ref{long_sen} we illustrate individual longitudinal profiles generated by 
SENECA (hybrid approach) code. Displayed cases correspond to median longitudinal 
profile produced for 1,000 gamma (dashed), iron nucleus (dot-dashed) and proton 
(solid) induced showers at 10$^{19}$~eV, at zenith angle $\theta$=$45^{\circ}$ and
free first interaction point. The shaded bands which follow each median line correspond 
to the upper limit of 68\% of confidence level. 

\begin{figure}[h]
\centerline{
\includegraphics[width=8.5cm]{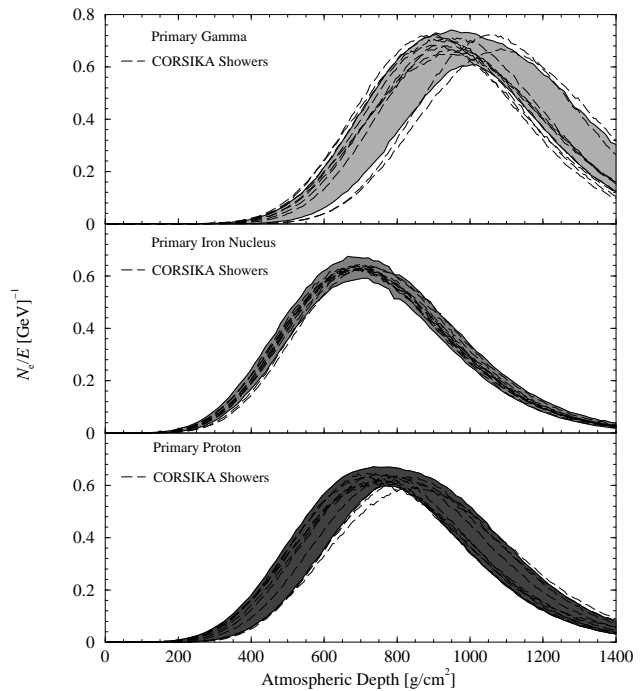}
}
\caption{Longitudinal profile of electrons-positrons for 10 random 
gamma, iron nucleus and proton initiated air showers at primary energy of 10$^{19}$~eV, 
with incident zenith angle of $45^{\circ}$, calculated by the CORSIKA (long-dashed) 
code, with the QGSJET01 hadronic interaction model. The shaded curve in each panel
represent the upper limit of 68\% of confindence level for 1,000 events simulated
with the SENECA code.}
\label{long_cors_sen}
\end{figure}

In order to make a simple comparison Fig.~\ref{long_cors_sen} illustrates the 
upper limit of 68\% of confindence level for 1,000 SENECA profiles with 10 
random showers simulated with CORSIKA. It is
possible to see a very reasonable agreement among the several longitudinal profiles.
In spite of that, we can verify that CORSIKA produces, apparently, more fluctuations
related to individual proton and gamma air showers developments. 
All simulations performed by the CORSIKA code were obtained by using the
thinning factor $t_{\mathrm{f}}$=10$^{-6}$ and the same 
energy threshold of SENECA simulations, 0.3~MeV (50~MeV) for photons and 
electrons (hadrons and muons). 


\begin{figure}[t]
\centerline{
\includegraphics[width=8.5cm]{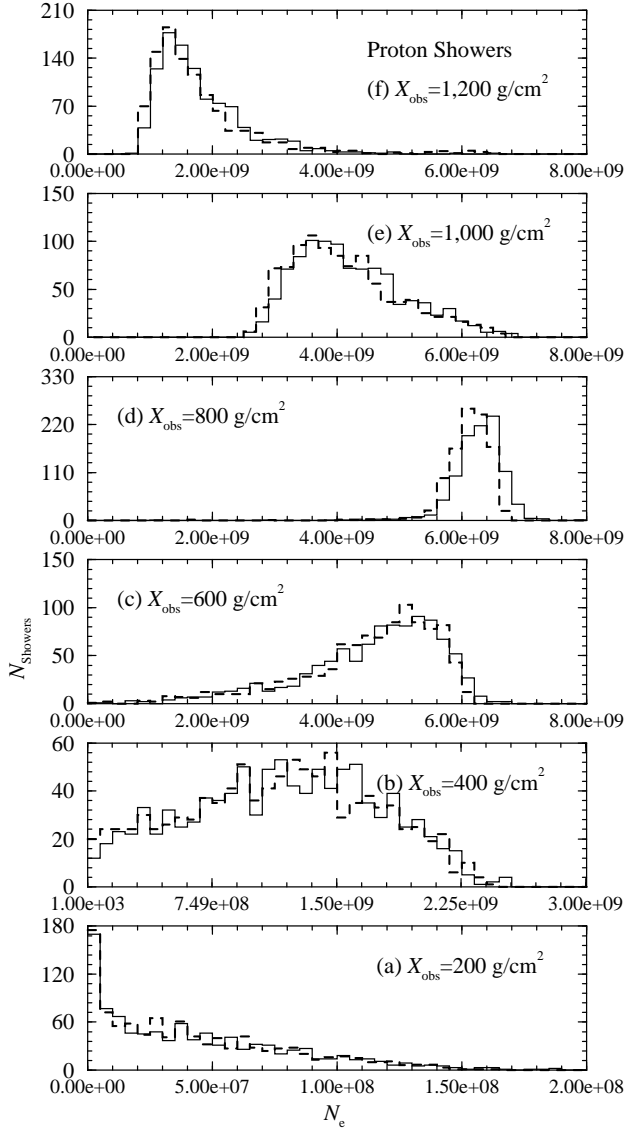}
}
\caption{Shower distribution in number of electrons at different slanth depths.
Results are shown for 1,000 showers, at 45$^{\circ}$, generated by primary
protons of energies $10^{19}$~eV calculated with the SENECA method (solid) and
CORSIKA code (dashed), both using the QGSJET01 hadronic interaction model. Each panel
represents a particular and arbitrary slanth depth: (a) 200, (b) 400, (c) 600,
(d) 800, (e) 1,000 and (f) 1,200, all in g/cm$^{2}$.}
\label{distr_el}
\end{figure}

To analyze more carefully the longitudinal profiles produced with the hybrid approach, 
we illustrate in Fig.~\ref{distr_el} the number of electrons-positrons generated by 1,000 proton initiated 
showers at energy of $10^{19}$~eV, with zenith angle 45$^{\circ}$, at arbitrary observation 
depths of (a) 200, (b) 400, (c) 600, (d) 800, (e) 1,000, and (f) 1,200 g/cm$^{2}$. 
We compare the hybrid results (solid line) to the results obtained with CORSIKA
(dashed line). The expected fluctuations due to very first proton interactions are
reflected on panel (a) and (b), producing a difference of $\sim$~4\% and
$\sim$~3\% between the distribution mean values. Panels (c) and (d) show 
discrepancies for the mean values of about 7\% and 6\%. In contrast to this,
the discrepancies basically disapear at panels (e) and (f), both obtaining 
an agreement of about 99.5\%. In all found discrepancies, SENECA produces more particles
than CORSIKA.  According to \cite{Drescher03}, some discrepancies are expected in the values produced 
for some shower quantities by the hybrid scheme and, apparently, 
are due to the combination of the used binning of discrete energy in the 
numerical solutions and the minimum energy threshold used for the 
electromagnetic cascade equations. 

Such differences between the results of both codes at the first
stages of the EAS development should not be crucial since depths around
and after the shower maximum are the most important ones for all fluorescence
experiments.

\subsection{Analyses of influence of input parameters\\on shower quantities}

In order to verify the dependence of the SENECA results, as suggested by the
SENECA authors, on the combination of used energy binning and minimum energy threshold 
($E_{\mathrm {min}}^{\mathrm{em}}$) we have simulated air showers induced by gamma
and proton primaries, with different initial conditions. Such primary particles 
have been chosen due to the different shower development in the atmosphere. 
The simulation development for gamma-induced showers deals mostly with the
electromagnetic processes while the simulation development for
proton (or any other hadron) showers deals with hadronic and
electromagnetic processes.
In other words, we are going to have different request of the cascade equation system 
for proton and gamma primary particles in the hybrid simulation. 

\subsubsection{$S_{\mathrm {max}}$ Parameter}

For the test with protons, 500 showers were generated of primary energy 
10$^{19}$~eV, with incident zenith angle of 45$^{\circ}$, calculated with the hybrid method 
using 30, 40 and 50 bins (10 bins) in the numerical solutions of electromagnetic 
(hadronic) cascades, and different minimum energy thresholds for the electromagnetic 
cascade equations $E_{\mathrm {min}}^{\mathrm{em}}$=~1, 10 and 100~GeV.
Following \cite{Drescher03}, we adopted $E_{\mathrm {min}}^{\mathrm{had}}$=~10$^{4}$~GeV. 
\begin{figure}[t]
\centerline{
\includegraphics[width=8.5cm]{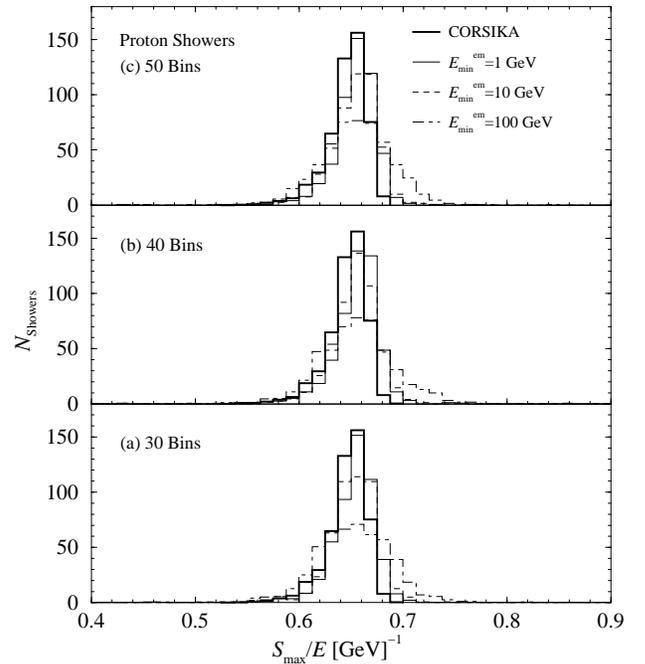}
}
\caption{Distribution of $S_{\mathrm {max}}$ normalized by the
primary energy in GeV.  Results are shown for 500 proton showers
of energy $10^{19}$~eV, with incident zenith angle of 45$^{\circ}$, calculated
with the hybrid scheme with different minimum energy thresholds and binnings of
discrete energy, for the electromagnetic cascade equations. The thick solid line
refers to the CORSIKA distribution. Both simulation codes generated the results
by using the QGSJET01 hadronic interaction model. Panels (a), (b) and (c) refer 
to the parameter values of 30, 40 and 50 bins of discrete energy, respectively, 
in the numerical solutions of the hybrid simulations.}
\label{thresh_distr_protons}
\end{figure}

\begin{table*}
\caption{Average values of $S_{\mathrm {max}}$ (standard deviation), normalized by the primary 
energy in GeV, obtained by hybrid simulations of 500 proton showers of primary energy 
$E=10^{19}$~eV, with incident zenith angle of 45$^{\circ}$. The predictions refer to the hybrid
scheme by using different minimum electromagnetic energy threshold and distint binnings of discrete
energy. The average value obtained with the CORSIKA code is
0.646 while the sigma is 2.06$\times$10$^{-2}$.}
\label{comparison}
\renewcommand{\arraystretch}{2.00}
\begin{tabular}{cc|cc|cc|cc} \hline \hline

& \multicolumn{3}{c}{30 Bins} & \multicolumn{2}{c}{40 Bins}& \multicolumn{2}{c}{50 Bins}\\  \hline


& & $S_{\mathrm {max}}/E$ & ($\sigma$) & $S_{\mathrm {max}}/E$ & ($\sigma$) & $S_{\mathrm {max}}/E$ & ($\sigma$)\\ \hline

$E_{\mathrm {min}}^{\mathrm{em}}$=1 GeV& &  0.653 & (2.08$\times$10$^{-2}$)& 0.655 & (2.75$\times$10$^{-2}$)& 0.653 & (2.85$\times$10$^{-2}$)\\ \hline\hline
$E_{\mathrm {min}}^{\mathrm{em}}$=10 GeV& &  0.651 & (2.79$\times$10$^{-2}$)& 0.653 & (2.61$\times$10$^{-2}$)& 0.654 & (2.55$\times$10$^{-2}$)\\ \hline\hline
$E_{\mathrm {min}}^{\mathrm{em}}$=100 GeV& &  0.651 & (4.27$\times$10$^{-2}$)& 0.654 & (4.05$\times$10$^{-2}$)& 0.653 & (4.09$\times$10$^{-2}$)\\ \hline\hline

\end{tabular}

\end{table*}


Fig.~\ref{thresh_distr_protons} shows the distribution of $S_{\mathrm {max}}$ normalized by the
primary energy in GeV, calculated with 500 primary protons by using 30 (panel a), 40 (panel b) 
and 50 (panel c) bins in the numerical solutions. The solid, dashed and dot-dashed lines
correspond to $E_{\mathrm {min}}^{\mathrm{em}}$=~1, 10 and 100~GeV, respectively, while
the thick solid line refers to the distribution obtained with the CORSIKA code.

Although we have modified the inputs for the SENECA simulations, the average 
values of $S_{\mathrm {max}}/\mathrm{E}$ do not vary significantly and are in total 
agreement (99\% on the average) with the average $S_{\mathrm {max}}/\mathrm{E}$ obtained 
with the CORSIKA shower generator. 
However, it is possible to verify in Fig.~\ref{thresh_distr_protons} that the $S_{\mathrm {max}}$
distributions for the hybrid code are visibly wider for the minimum energy 
threshold values of 10 and 100~GeV, when compared to CORSIKA, showing a different 
behaviour for the $S_{\mathrm {max}}$ fluctuations in both codes. Besides, such fluctuation description 
changes dramatically the characteristic asymmetrical shape of the $S_{\mathrm {max}}$ distribution.
The average (fluctuation) values of $S_{\mathrm {max}}/E$ calculated with the hybrid scheme
are presented in Table~\ref{comparison} and correspond to the distributions illustrated in 
Fig.~\ref{thresh_distr_protons}. The average value obtained with the CORSIKA code is 
0.646 while the width distribution is 2.06$\times$10$^{-2}$. 

\begin{figure}[t]
\centerline{
\includegraphics[width=8.5cm]{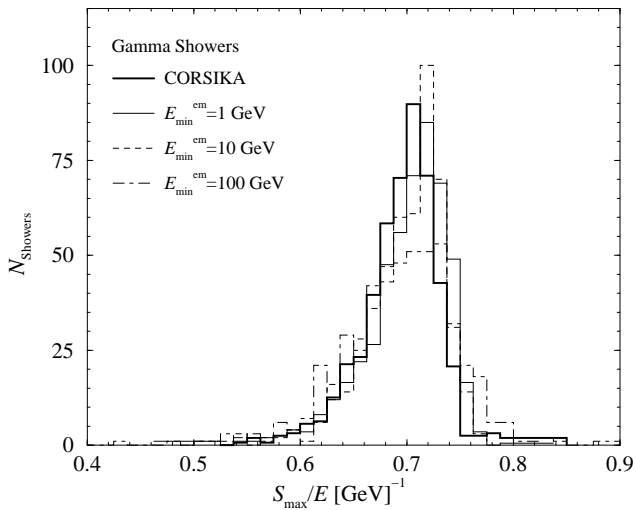}
}
\caption{Distribution of $S_{\mathrm {max}}$ normalized by the
primary energy in GeV.  Results are shown for 500 primary gamma showers
of energy $10^{19}$~eV, with zenith angle of 45$^{\circ}$ calculated
with the SENECA and CORSIKA (thick solid line) schemes by using the QGSJET01 
hadronic interaction model. The SENECA simulations were performed with 
different minimum energy thresholds for the electromagnetic cascade equations 
and the input value of 30 bins in the numerical solutions.}
\label{thresh_distr_gammas}
\end{figure}

Such $S_{\mathrm {max}}$ fluctuation descriptions for the 
minimum electromagnetic energy threshold values of 10 and 100~GeV can clearly be seen as very strong dependence 
of shower quantities on initial parameters and are much larger when compared
to the standard deviation value predicted by the CORSIKA code.
One equally important aspect is that we have verified the dependence of quantities on
higher values of $E_{\mathrm {min}}^{\mathrm{em}}$, while \cite{Drescher03}
expects the dependences on lower $E_{\mathrm {min}}^{\mathrm{em}}$ values. 

The same unsual fluctuation description for $S_{\mathrm {max}}$ can also be visualized 
in Fig.~\ref{thresh_distr_gammas}, which shows the $S_{\mathrm {max}}/\mathrm{E}$ 
distribution calculated with primary gammas by using 30 bins as input value in 
the numerical solutions. The thick solid line refers to the distribution obtained 
with the CORSIKA code. The average (sigma) values obtained from SENECA are 
0.7 (4.$\times$10$^{-2}$), 0.698 (4.02$\times$10$^{-2}$) and 0.694 (7.79$\times$10$^{-2}$), 
for $E_{\mathrm {min}}^{\mathrm{em}}$=~1, 10 and 100~GeV, respectively. 
The CORSIKA predicts the average $S_{\mathrm {max}}/\mathrm{E}$ distribution 
value of 0.694 and the width distribution of 4.74$\times$10$^{-2}$.

Hence, these checks basically verify that, apparently, the minimum electromagnetic 
energy thresholds of 10 and 100~GeV may produce artificial fluctuations on 
$S_{\mathrm {max}}$.  Moreover, such fluctuation description can introduce 
statistical errors on analyses of event-to-event and could be possible critics in 
pratical applications of SENECA in fluorescence event reconstruction.

The increase in fluctuations with increasing $E_{\mathrm {min}}^{\mathrm{em}}$, is
difficult to understand since the larger the $E_{\mathrm {min}}^{\mathrm{em}}$,
the greater the weighted of the Monte Carlo portion of the hybrid scheme.

\begin{figure}[t]
\centerline{
\includegraphics[width=8.5cm]{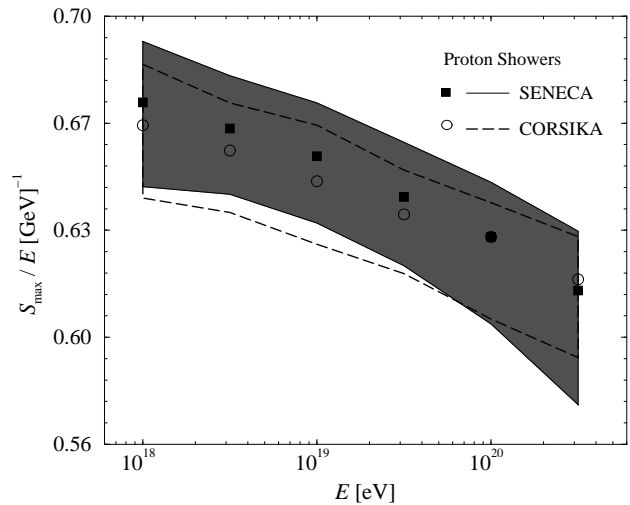}
}
\caption{Median values of the shower size at maximum air shower development
($S_{\mathrm {max}}$), normalized by the primary energy in GeV, related to 
proton-induced showers as a function of primary energy. The solid 
(long-dashed) line represents a band containing 68\% of confindence 
level for 500 (100) events generated by the SENECA
(CORSIKA) code, at $\theta=45^\circ$, using the QGSJET01 hadronic
interaction model. The full square (full circle) represents the median
values of $S_{\mathrm {max}}$ obtained with the SENECA (CORSIKA) code.}
\label{smax_ave}
\end{figure}

Fig.~\ref{smax_ave} shows the median shower size values, normalized by the
primary energy in GeV, at the depth of shower maximum. Each band contains
68\% of confidence level and is related to proton initiated
showers as a function of primary energy ($E$$>$10$^{18}$~eV).
The solid (dashed) line illustrates the predictions of the SENECA technique
(CORSIKA), using the QGSJET01 model. We have used $E_{\mathrm {min}}^{\mathrm{em}}$=~1~GeV
and 30 bins of discrete energy as the input SENECA parameters. 

The values predicted by both codes have a visible agreement ($>$~99\%). 
SENECA tends to produce systematically higher values of $S_{\mathrm {max}}$ than 
CORSIKA, up to 10$^{19.5}$~eV, and lower values for energies $>$10$^{20}$~eV. 
Also, it seems that SENECA produces more fluctuations related to
smaller shower sizes at energies around 10$^{20}$~eV.

\subsubsection{$X_{\mathrm {max}}$ Parameter}

As discussed in Sec.~\ref{introduction}, fluorescence detectors observe
the longitudinal profile evolution by measuring the fluorescence light.
The greatest advantage of this technique is that it allows the estimation of the number
of charged particles as a function of depth in the atmosphere, from the measured
\begin{figure}[t]
\centerline{
\includegraphics[width=8.5cm]{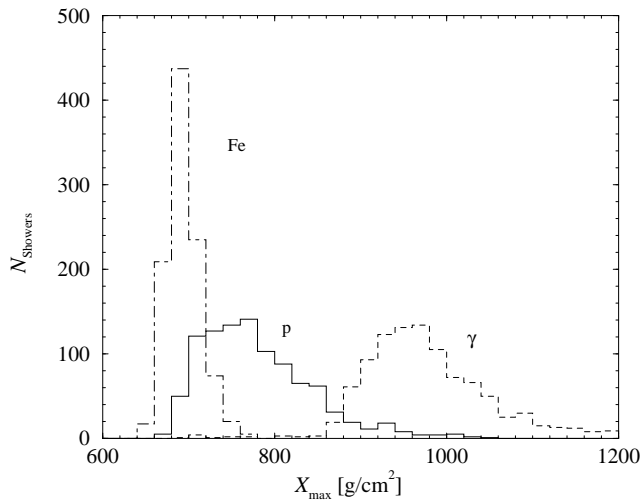}
}
\caption{Distribution of the depth of maximum air shower development shown 
for 1,000 proton, iron nucleus and gamma showers at a particular primary 
energy of $10^{19}$~eV, with incident zenith angle of 45$^{\circ}$,
calculated with the hybrid technique.}
\label{xmax_distri_ga_fe_pr}
\end{figure}
data~\cite{Song00,Baltrusaitis85,Barbosa03,Alvarez04}, which makes possible, in turn,
the estimation of the depth of maximum development of the shower, $X_{\mathrm {max}}$.

\begin{figure}[b]
\centerline{
\includegraphics[width=8.5cm]{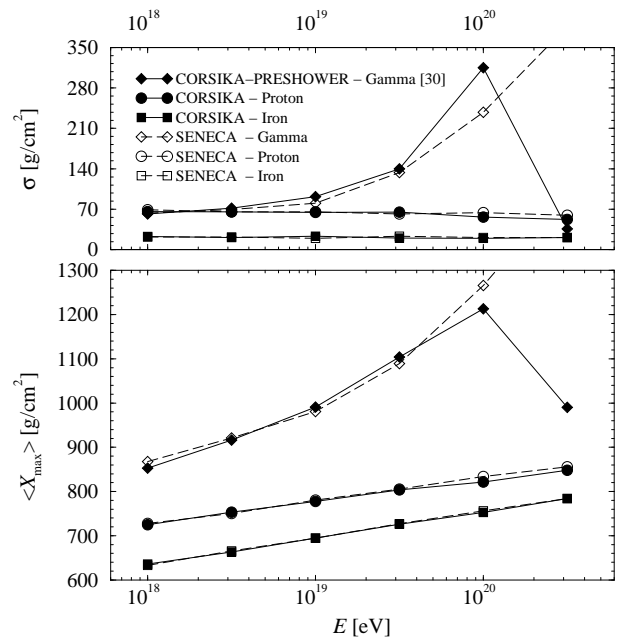}
}
\caption{Average depth (bottom panel) and sigma (top panel) values of maximum air shower development
($X_{\mathrm {max}}$) related to gamma, iron nucleus and proton induced 
showers as a function of extremely high primary shower energy, at 
$\theta=45^\circ$, using the QGSJET01 hadronic interaction model. The 
long-dashed lines represent 500 events per energy induced by gamma (diamond),
iron nucleus (square) and proton (circle) primaries generated 
by the SENECA. The CORSIKA predictions correspond to 100 simulated showers
and are represented by solid lines and full symbols~\cite{Heck04}.}
\label{xmax_ave}
\end{figure}
\begin{figure}[h]
\centerline{
\includegraphics[width=8.5cm]{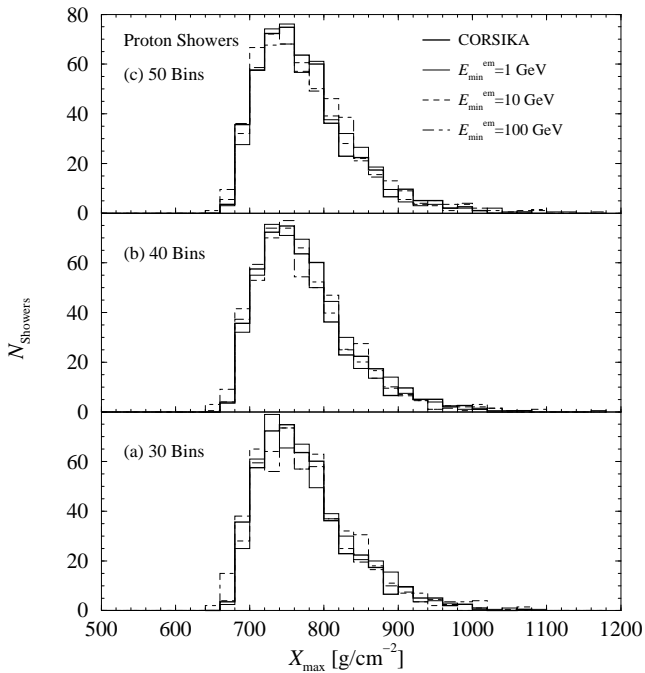}
}
\caption{Distribution of the depth of maximum air shower development. Results 
are shown for 500 proton showers of energy $10^{19}$~eV, with 45$^{\circ}$, 
calculated with the SENECA by using different minimum energy thresholds and 
binnings of discrete energy, for the electromagnetic 
cascade equations. The thick solid line refers to the CORSIKA distribution. 
Both simulation codes generated the results by using the QGSJET01 hadronic 
interaction model. Panels (a), (b) and (c) refer to the parameter values of 30, 
40 and 50 bins of discrete energy, respectively, in the numerical solutions 
of the hybrid simulations.}
\label{xmax_distr_protons}
\end{figure}
In principle, obtaining the values of $X_{\mathrm {max}}$ and/or $S_{\mathrm {max}}$,
by the fluorescence technique, and their respective fluctuations, by Monte Carlo, one 
should be able to reconstruct the shower energy and infer the identity of the primary 
cosmic ray~\cite{Alvarez04,Cronin95}. Fig.\ref{xmax_distri_ga_fe_pr} illustrates the
potential of the $X_{\mathrm {max}}$ distribution, generated with
the hybrid scheme, to distinguish possible primary signatures, as calculated using
SENECA.

We have also checked the behaviour of average $X_{\mathrm{max}}$ values for gamma,
iron nucleus and proton induced air shower.

Fig.~\ref{xmax_ave} shows the average depth (bottom panel) and fluctuation 
(top panel) values of shower 
maximum ($X_{\mathrm {max}}$) for the energy range of 10$^{18}$-10$^{20.5}$~eV,
with incident zenith angle of $\theta=45^\circ$. All simulations performed
in this plot were generated with the QGSJET01 hadronic interaction model.
The long-dashed line represents 500 showers induced by gamma (diamond),
iron nucleus (square) and proton (circle) primaries at each energy generated
by SENECA. The CORSIKA predictions correspond to 100 simulated showers and 
are represented by solid lines and full symbols \cite{Heck04}. Only in Fig.~\ref{xmax_ave}
the CORSIKA results obtained for gamma showers were generated by using the
PRESHOWER code. All other CORSIKA simulations in the present work were calculated
without the PRESHOWER code.
\begin{figure}[h]
\centerline{
\includegraphics[width=8.5cm]{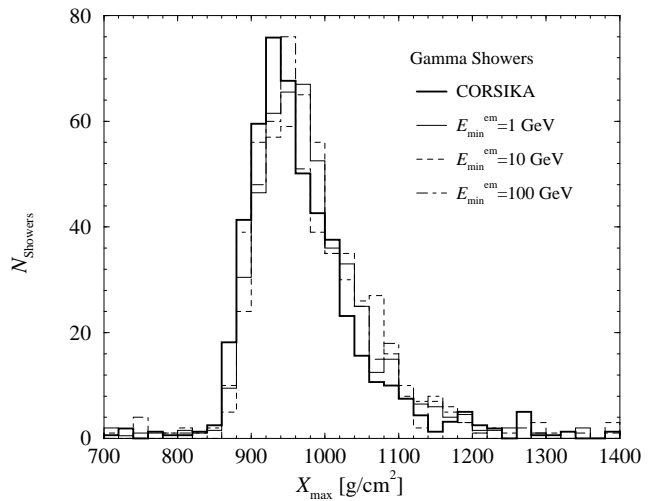}
}
\caption{Distribution of the depth of maximum air shower development. Results 
are shown for 500 gamma showers of energy $10^{19}$~eV, with incident zenith 
angle of 45$^{\circ}$, calculated by the hybrid scheme with different minimum 
energy thresholds for the electromagnetic 
cascade equations. The thick solid line refers to the CORSIKA distribution. 
The obtained values refer to 30 bins of discrete energy
in the numerical solutions of the hybrid simulations.}
\label{xmax_distr_gammas}
\end{figure}

The values predicted by both codes are in very good agreement, showing 
almost identical values for $X_{\mathrm {max}}$ and fluctuations,
for all primary particles. One interesting aspect to be mentioned is the strong influence of
the geomagnetic field on depth of maximum development in gamma showers, 
generated by CORSIKA code~\cite{Heck04} implemented with the PRESHOWER program~\cite{Homola04}, 
at energies above 10$^{20}$~eV. The gomagnetic field decelerates the air
gamma shower development, modifying the depth in which occurs the maximum
development. The average (fluctuation) value of $X_{\mathrm {max}}$ decreases from 1,213 (314) g/cm$^{2}$
at 10$^{20}$~eV to 990 (35) g/cm$^{2}$ at 10$^{20.5}$~eV.
Altough we have considered the geomagnetic field in our simulations, it
seems that SENECA do not consider the interaction of ultra high energy
photons with the Earth's geomagnetic field  before entering the Earth's
atmosphere. Such effect absence produces the discrepancy
seen in gamma curves at energies above and around 10$^{20}$~eV. References
\cite{Homola04b,Risse04c} discuss very high energy gamma showers. 

Fig.~\ref{xmax_distr_protons} confirms the strength of the agreement by showing the 
distribution of $X_{\mathrm {max}}$ values, produced by 500 proton showers, for 
the particular primary energy $10^{19}$~eV, with incident zenith angle of 45$^{\circ}$, 
calculated with SENECA and CORSIKA (thick solid line) codes and QGSJET01 
hadronic interaction model. We used the same input parameters
for SENECA as in Fig.~\ref{thresh_distr_protons}.
The mean (fluctuation) $X_{\mathrm {max}}$ values calculated with the hybrid scheme
are presented in Table~\ref{comparison_xmax}. 
These values are in total agreement with CORSIKA values, 
$\langle X_{\mathrm {max}}\rangle$=774~g/cm$^{2}$ 
($\sigma$=65~g/cm$^{2}$), and show, apparently, that the influence of minimum energy
threshold and binning of discrete energy on $X_{\mathrm {max}}$ values
is negligible in hadron initiated showers.
\begin{table*}
\caption{Average (standard deviation) values of $X_{\mathrm {max}}$, all in g/cm$^{2}$, obtained 
by hybrid simulations of 500 proton showers of primary energy $E=10^{19}$~eV, 
with incident zenith angle of 45$^{\circ}$. The predictions refer to the hybrid
scheme by using different minimum electromagnetic energy threshold and distint 
binnings of discrete energy. The average value obtained with the CORSIKA code is
774~g/cm$^{2}$ while the sigma is 65~g/cm$^{2}$.}
\label{comparison_xmax}
\renewcommand{\arraystretch}{2.00}
\begin{tabular}{cc|cc|cc|cc} \hline \hline

& \multicolumn{3}{c}{30 Bins} & \multicolumn{2}{c}{40 Bins}& \multicolumn{2}{c}{50 Bins}\\  \hline

& & $X_{\mathrm {max}}$ & ($\sigma$) & $X_{\mathrm {max}}$ & ($\sigma$) & $X_{\mathrm {max}}$ & ($\sigma$)\\ \hline

$E_{\mathrm {min}}^{\mathrm{em}}$=1 GeV& &  778 & (65)& 775 & (63)& 776 & (64)\\ \hline\hline
$E_{\mathrm {min}}^{\mathrm{em}}$=10 GeV& &  775 & (62)& 772 & (61)& 774 & (65)\\ \hline\hline
$E_{\mathrm {min}}^{\mathrm{em}}$=100 GeV& &  774 & (69)& 771 & (66)& 777 & (70)\\ \hline\hline

\end{tabular}
\end{table*}

We have made the same $X_{\mathrm {max}}$ verification for gamma showers in 
Fig.~\ref{xmax_distr_gammas} and we verify that CORSIKA and SENECA
distributions are in good agreement. SENECA predicts the average 
(fluctuation) $X_{\mathrm {max}}$ distribution values of 981~g/cm$^{2}$
(81~g/cm$^{2}$), 989~g/cm$^{2}$ (85~g/cm$^{2}$) and 985~g/cm$^{2}$ (101~g/cm$^{2}$), 
for minimum electromagnetic energy thresholds of 1, 10 and 100~GeV, respectively.
CORSIKA code, without PRESHOWER, produces the average and width distribution values of 976~g/cm$^{2}$
and 97~g/cm$^{2}$, respectively. Due to the extreme fluctuations in shower development induced by
primary gammas, which affect directly the position of shower maximum development,
the average and fluctuation results obtained by SENECA seem to be 
reasonable, in spite of the fact that again the fluctuation values increase with
the electromagnetic minimum energy thresholds of 10 and 100~GeV, over those of CORSIKA.


\subsubsection{$S_{\mathrm {max}}$-$X_{\mathrm {max}}$ Correlation}

We also analyzed the correlation between $S_{\mathrm {max}}$ and
$X_{\mathrm {max}}$, which are important longitudinal shower
quantities on event reconstruction. We compared SENECA results with 
CORSIKA predictions for gamma and proton induced air shower.

\begin{figure}[h]
\centerline{
\includegraphics[width=8.5cm]{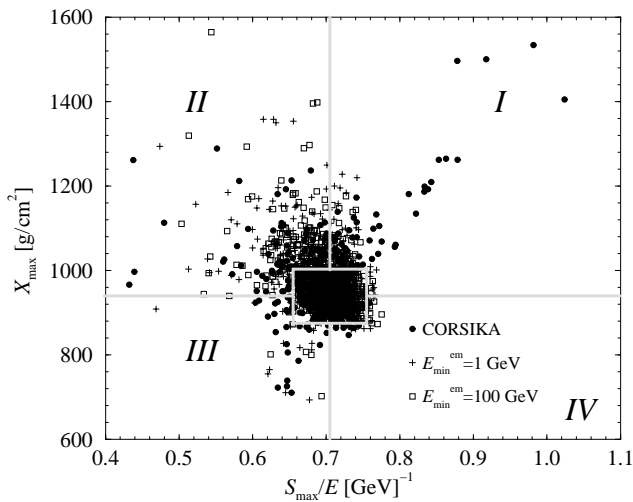}
}
\caption{The correlation between $X_{\mathrm {max}}$ and $S_{\mathrm {max}}/$
for gamma induced showers at primary energy $10^{19}$~eV, at zenith angle
$\theta=45^\circ$, obtained with SENECA and CORSIKA codes. The full circles represent
500 showers simulated with CORSIKA, while square and cross symbols illustrate 500 events
generated with the hybrid method, by using minimum electromagnetic energy thresholds
$E_{\mathrm {min}}^{\mathrm{em}}$=~1 and 100~GeV, respectively.}
\label{smax_xmax_gamma}
\end{figure}

Fig.~\ref{smax_xmax_gamma} shows the correlation between these parameters
for both simulation schemes. We simulated 500 gamma showers for each code,
of energy 10$^{19}$~eV, incident zenith angle $\theta$=45$^{\circ}$ and
free first interaction point.
Hybrid results are shown for minimum energy thresholds
$E_{\mathrm {min}}^{\mathrm{em}}$=~1 (crosses) and 100~GeV (squares), and
30 bins of discrete energy in the numerical solutions. The full circles
denote the corresponding values for CORSIKA. It is possible to verify
the existence of large fluctuations in gamma air showers at this particular
primary energy.

The small box in the figure encloses, approximately, the highest density of
correlation points for both codes: 61\% and 59\% of the total number of events
simulated by SENECA, for $E_{\mathrm {min}}^{\mathrm{em}}$=~1 and 100~GeV,
respectively, and 64\% of the total showers generated with CORSIKA.

\begin{figure}[h]
\centerline{
\includegraphics[width=8.5cm]{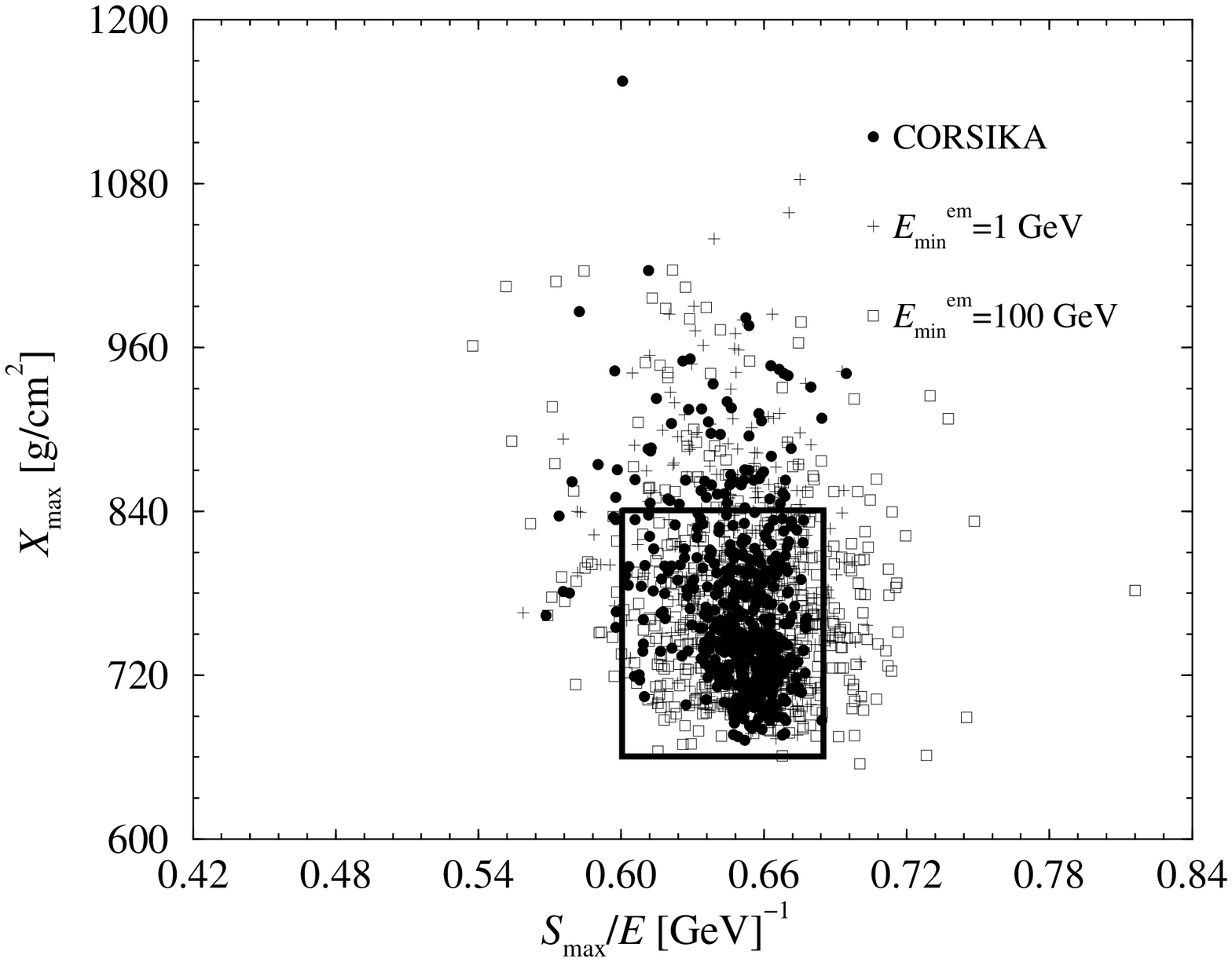}
}
\caption{The correlation between $X_{\mathrm {max}}$ and $S_{\mathrm {max}}/E$
for proton induced showers at primary energy $10^{19}$~eV, at zenith angle
$\theta=45^\circ$, obtained with SENECA and CORSIKA codes. The full circles represent
500 showers simulated with CORSIKA, while square and cross symbols correspond to 500 showers
generated with the hybrid method, by using minimum electromagnetic energy thresholds
$E_{\mathrm {min}}^{\mathrm{em}}$=~1 and 100~GeV, respectively.}
\label{smax_xmax_proton}
\end{figure}

\begin{figure}[h]
\centerline{
\includegraphics[width=8.5cm]{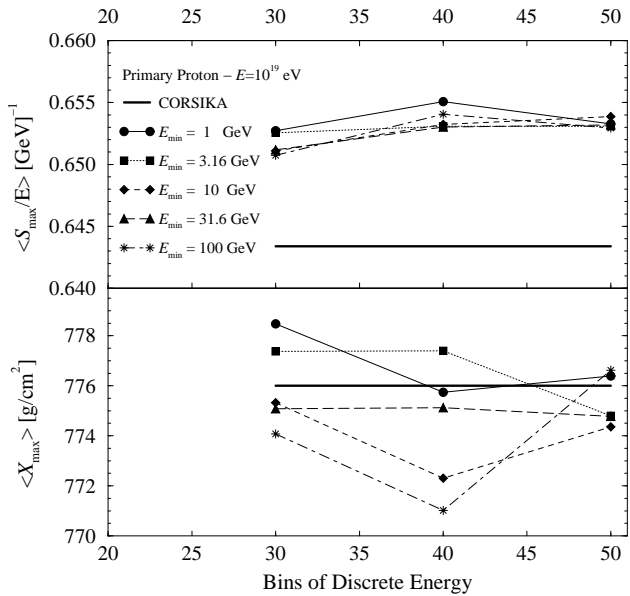}
}
\caption{Average values of $X_{\mathrm {max}}$ and $S_{\mathrm {max}}/$
for 1,000 proton induced showers at primary energy $10^{19}$~eV, at zenith angle
$\theta=45^\circ$, obtained with SENECA and CORSIKA (thick solid line) codes (see the text).}
\label{smax_xmax_meanvalues}
\end{figure}

The number of gamma showers predicted by SENECA and CORSIKA vary considerably
from one quadrant to another in Fig.~\ref{smax_xmax_gamma}, indicating
qualitative and quantitative differences in the character of the fluctuations in
both codes.

The fractions of events inside regions I, II, III and IV (see the figure) are, 
respectively, 10\% (10\%), 24\% (27\%), 3\% (2\%) and 2\% (2\%), for the minimum energy threshold of 1
(100)~GeV. In the same regions, the corresponding CORSIKA fractions are 11\%, 17\%, 6\% and 2\%.

Notoriously, CORSIKA shows in region I  a particular structure related to maximum depth development
($X_{\mathrm{max}}$$>$940~g/cm$^{2}$) of particularly large showers ($S\mathrm{max}$/$E>$0.705). 
Such structure is absent in SENECA.

On the other hand, the hybrid scheme predicts a significantly number of ordinary and small showers
in region II (118 and 137 showers for $E_{\mathrm {min}}^{\mathrm{em}}$=~1 and 
100~GeV, respectively), which achieve higher values for the depth of maximum 
development, a behavior that is not as strong in CORSIKA (84 showers).

The same kind of comparison is presented in Fig.~\ref{smax_xmax_proton} for proton showers.
The same input parameters as in the previous figure were used to simulate the
three sets of 500 proton showers each displayed in Fig.~\ref{smax_xmax_gamma}.

It is apparent from the figure that the hybrid approach generates a more scattered
distribution of showers, with wider tales, than CORSIKA does.

This is confirmed by counting the number of events inside the
small box shown in the figure. SENECA expectation amounts to 80\% (69\%)
of the proton showers falling inside the box for $E_{\mathrm {min}}^{\mathrm{em}}$=~1 and
100~GeV, while CORSIKA expectation is 85\%.
\begin{figure}[t]
\centerline{
\includegraphics[width=8.5cm]{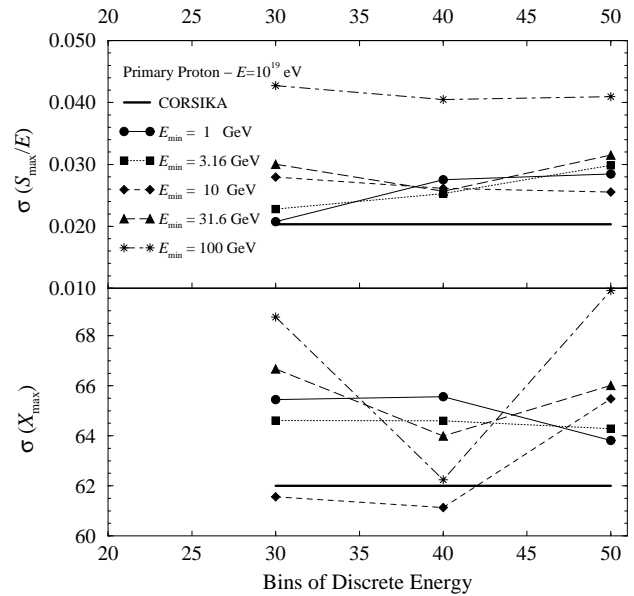}
}
\caption{Fluctuation of the position of shower maximum and maximum shower size,
for 1,000 proton induced showers at primary energy $10^{19}$~eV, at zenith angle
$\theta=45^\circ$, obtained with SENECA and CORSIKA (thick solid line) codes (see the text).}
\label{smax_xmax_sigmavalues}
\end{figure}

We can verify in SENECA predictions that the input parameter
$E_{\mathrm {min}}^{\mathrm{em}}$=100~GeV produces larger fluctuations 
related to proton shower size ($\sigma_{S\mathrm{max}/E}$), describing a sort of symetrical distribution
of events outside the box. The standard deviation of proton shower size,
obtained for this particular input parameter in SENECA, is 3.62$\times$10$^{-2}$
while the corresponding value for CORSIKA is 1.89$\times$10$^{-2}$. Such symetrical 
statement is confirmed in Fig.~\ref{thresh_distr_protons}.

Figs.~\ref{smax_xmax_meanvalues} and \ref{smax_xmax_sigmavalues} illustrate
the dependence of the average values, and their fluctuation, of
$X_{\mathrm {max}}$ and $S_{\mathrm {max}}$ on input parameters of electromagnetic cascade equations
solved in SENECA scheme. We have generated 1,000 proton showers, at primary energy
$10^{19}$~eV and zenith angle $\theta=45^\circ$, using 30, 40 and 50 bins of possible
discrete energy for each minimum energy threshold; total of 15,000 simulated events.
As it has been noted in top panel of Fig.~\ref{smax_xmax_meanvalues}, the average values
produced by SENECA are in perfect agreement ($>$99\% for both quantities)
with the corresponding results obtained from CORSIKA (thick solid line).

However, Fig.~\ref{smax_xmax_sigmavalues} shows that the fluctuations related to $S_{\mathrm {max}}$ (top panel) have large discrepancies for a minimum electromagnetic energy threshold of 100~GeV. In this case, there is more than a factor of 2 between SENECA and CORSIKA expected fluctuation levels ($\sigma \sim 4.15 \times 10^{-2}$, averaged over number of bins, vs. $\sigma \sim 2.034 \times 10^{-2}$ in the case of CORSIKA).

On average, the discrepancies in $\sigma$ between SENECA and CORSIKA decrease systematically at progressively lower values of $E_{\mathrm {min}}^{\mathrm{em}}$.

Even if fluctuations in $X_{\mathrm {max}}$ can be up to 8\% larger in SENECA than in CORSIKA (for  $E_{\mathrm {min}}^{\mathrm{em}}$=100GeV) this discrepancy
is of small practical relevance when translated into units of depth (i.e., in g/cm$^{2}$).


\subsubsection{$N_{\mu}$ - Number of muons}

To ensure the consistency of other longitudinal shower components in SENECA,
we have also compared the number of muons produced by both SENECA and CORSIKA
codes.

\begin{figure}[h]
\centerline{
\includegraphics[width=8.5cm]{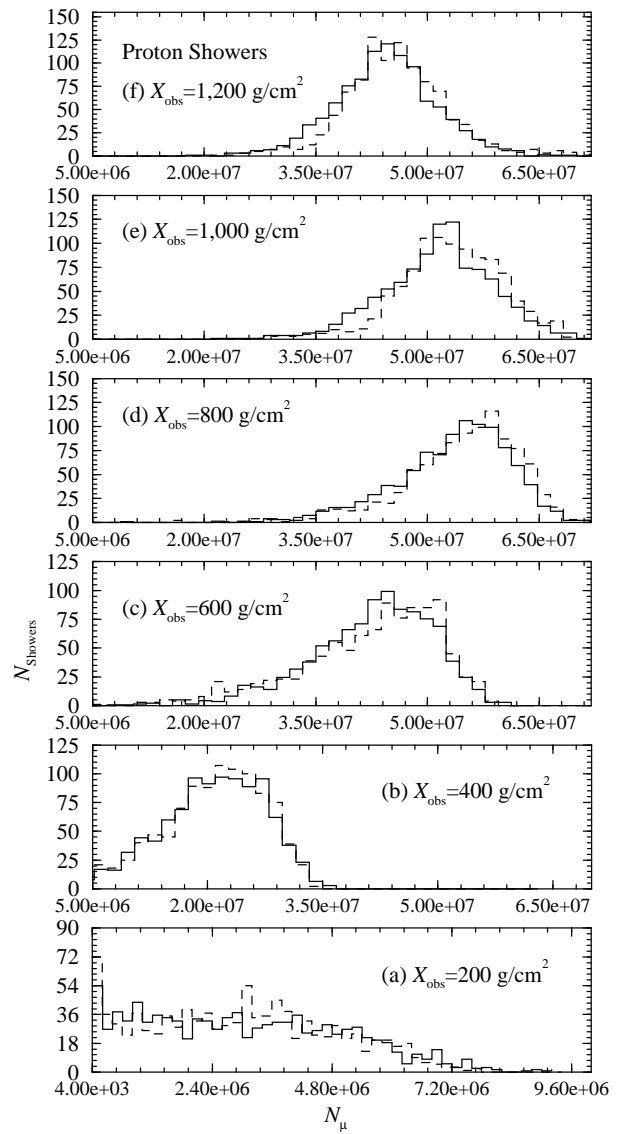}
}
\caption{Shower distribution in number of muons at different slanth depths.
Results are shown for 1,000 showers, at 45$^{\circ}$, generated by primary
protons of energies $10^{19}$~eV calculated with the SENECA method (solid) and
CORSIKA code (dashed), both using the QGSJET01 hadronic interaction model. Each panel
represents a particular and arbitrary slanth depth: (a) 200, (b) 400, (c) 600,
(d) 800, (e) 1,000 and (f) 1,200, all in g/cm$^{2}$.}
\label{distr_mu}
\end{figure}

Fig.~\ref{distr_mu} shows the muon component for 1,000 proton initiated showers at energy of
$10^{19}$~eV and incident zenith angle of 45$^{\circ}$, at observation depths of (a) 200,
(b) 400, (c) 600, (d) 800, (e) 1,000, and (f) 1,200 g/cm$^{2}$. As we can seen,
both distribution agree very well.
The average number of muons are very similar ($~$99\% of coincidence) when related to the first stages of
shower development, panels (a), (b) and (c). At the same time, discrepancies become
greater with depths closer to the sea level, but are on average less than $\sim 3$\% (2\% at 800~g/cm$^{2}$,
3.6\% at 1,000~g/cm$^{2}$ and 3.3\% at 1,200~g/cm$^{2}$). Such small discrepancies
can be considered negligible, because they are undetectable to any experiment
which measures the correlated number of particles. Furthermore, the fluctuations
(width distribution) corresponding to each panel agree well with those predicted
in CORSIKA code. Consequently, the muon distributions seem to be well described in
the hybrid scheme when compared to the CORSIKA ones.

\subsubsection{Processing time consumption}

As a final comparison, we show in Table~\ref{Table0} the impressive numbers related
to the rate of time-consumption ($\mathrm{T}_{\mathrm{CORS}}/\mathrm{T}_{\mathrm{SEN}}$) 
between the CORSIKA and SENECA codes. All simulations generated with CORSIKA were obtained by using the thinning
factor $t_{\mathrm{f}}$$=$10$^{-6}$. 

In order to compare the time-consuming of the hybrid method, by using the thinning procedure 
in the Monte Carlo calculation, we have generated air showers with thinning level of 10$^{-6}$ 
for the Monte Carlo scheme.

In Table~\ref{Table0}, Rate-1 refers to the hybrid simulations without the thinning procedure 
while Rate-2 considers the SENECA simulations by using the thinning procedure with thinning 
level of 10$^{-6}$ in the Monte Carlo scheme.

\begin{table*}
\renewcommand{\arraystretch}{1.5}
\caption{\label{Table0} Rate of time-consuming between CORSIKA and SENECA codes, i.e.,
Rate=$\mathrm{T}_{\mathrm{CORS}}/\mathrm{T}_{\mathrm{SEN}}$. Rate-1 refers to the
hybrid scheme with the traditional Monte Carlo method, while
Rate-2 considers the hybrid scheme using the thinning level of 10$^{-6}$ in the Monte
Carlo. All CPU times used for the calculation of the rate refer to a 2.1 GHz AMD Athlon processor.}
\vspace{0.1cm}
\begin{tabular}{ccc} \hline\hline
$\log_{10} E_{0}$~[eV] & Rate-1 & Rate-2\\ \hline
 17.0 &  3  & 13 \\
 17.5 &  3  & 13 \\
 18.0 &  4  & 15 \\
 18.5 &  5  & 16 \\
 19.0 &  7  & 20 \\
 19.5 &  11 &  28 \\
 20.0 &  20 &  46 \\
 20.5 &  35 & 74 \\ \hline\hline
\end{tabular}\\
\end{table*}

\section{\label{secIV}Discussion and conclusions}

In the present work we analyzed the practical potential of SENECA, a very fast hybrid 
tri-dimensional code, for the simulation of the longitudinal development of extensive 
air showers at high energies. We take as reference the well known and extensively tested 
CORSIKA code, which is based on a much slower, but highly reliable, full Monte Carlo 
simulation. The QGSJET01 hadronic interaction model is used throughout the paper for both codes.

Although a careful analysis of many shower quantities discussed here is strongly model 
dependent, we are confident that the present study is able to show the
potential and limitations of SENECA, in its present version, for a practical application 
to the analysis of fluorescence data on ultra-high energy cosmic rays.

The consistency of the SENECA scheme was tested and it proved to be very stable for energies 
above $10^{18}$ eV. The results obtained by both codes agree well with negligible discrepancies 
for most quantities analyzed.

Apparently, minimum electromagnetic energy threshold values of 10 and 100~GeV may produce 
artificial fluctuations on $S_{\mathrm {max}}$ shower quantity.  These could be responsible 
for the introduction of systematic errors on the analysis of isolated events and should be 
taken into account when applying SENECA to fluorescence reconstruction.

In any case, the undisputable bounty of SENECA is velocity. The rates of processing 
time-consumption between CORSIKA and SENECA are, to say the least, impressive over the wide 
range of primary shower energy tested here.

Finally, as a word of caution, special attention must be paid to the selection of the several 
input parameters of SENECA in practical applications, since they may affect the results in a 
non-trivial and somehow unpredictable way.

\noindent

{\bf Acknowledgments}
The authors acknowledge Hans-Joachim Drescher and Glennys Farrar                                                        for making SENECA code available to the cosmic ray academic community. We
are indebted to Dieter Heck for providing us with the values of                                                         $X_{\mathrm{max}}$ for CORSIKA.
J.A.Ortiz is supported by CNPq/Brazil, G.M.T. by FAPESP and CNPq and L.V. Souza by
FAPESP. Most of the simulations presented here were carried on a Cluster
Linux TDI (32 Dual Xeon 2.8 GHz, 2GBytes RAM nodes), supported by
Laborat\'orio de Computa\c c\~ao Cient\'{\i}fica Avan\c cada at
Universidade de S\~ao Paulo.

\end{document}